\documentclass[twocolumn,showpacs,preprintnumbers,amsmath,amssymb,superscriptaddress]{revtex4}

\usepackage{graphicx}
\usepackage{dcolumn}
\usepackage{bm}

\begin{document}

\title{Astrophysical S-factor of
$^{14}\rm N(\rm p,\gamma)^{15}\rm O$} 
\thanks{Supported in part by INFN, BMBF(05CL1PC1),
GSI(BO-ROL), OTKA(T034259 and T042733), TARI HPRI-CT-2001-00149 and SSTC
PAI(P5/07)}

\author{A.Formicola}
 \affiliation{Institut f\"ur Experimentalphysik III,
	Ruhr-Universit\"at Bochum, Bochum, Germany}
\author{G.Imbriani}
 \affiliation{Osservatorio Astronomico di Collurania, Teramo and INFN Napoli,
 	Italy}
\author{H.Costantini}
 \affiliation{Universit\`a di Genova, Dipartimento di Fisica and INFN Genova,
 	Italy}
\author{C.Angulo}
 \affiliation{Centre de Recherches du Cyclotron,
 	Universit\'e catholique de Louvain, Belgium}
\author{D.Bemmerer}
 \affiliation{Institut f\"ur Atomare Physik und Fachdidaktik,
 	Technische-Universit\"at Berlin, Germany}
\author{R.Bonetti}
 \affiliation{Universit\`a di Milano, Istituto di Fisica and INFN Milano, Italy}
\author{C.Broggini}
 \affiliation{INFN Padova, Italy }
\author{P.Corvisiero}
 \affiliation{Universit\`a di Genova, Dipartimento di Fisica and INFN Genova,
 	Italy}
\author{J.Cruz}
 \affiliation{Centro de Fisica Nuclear da Universidade de Lisboa, Lisboa, Portugal}
\author{P.Descouvemont}
 \affiliation{Physique Nucl\'eaire Th\'eorique and Physique Math\'ematique, CP229,\\
	Universit\'e Libre de Bruxelles, Brussels, Belgium}
\author{Z.F\"{u}l\"{o}p}
 \affiliation{ATOMKI, Debrecen, Hungary}
\author{G.Gervino}
 \affiliation{Universit\`a di Torino, Dipartimento di Fisica
	Sperimentale and INFN Torino, Italy}
\author{A.Guglielmetti}
 \affiliation{Universit\`a di Milano, Istituto di Fisica and INFN Milano, Italy}
\author{C.Gustavino}
 \affiliation{INFN Laboratori Nazionali del Gran Sasso, Assergi, Italy}
\author{G.Gy\"urky}
 \affiliation{ATOMKI, Debrecen, Hungary}
\author{A.P.Jesus}
 \affiliation{Centro de Fisica Nuclear da Universidade de Lisboa, Lisboa, Portugal}
\author{M.Junker}
 \affiliation{INFN Laboratori Nazionali del Gran Sasso, Assergi, Italy}
\author{A.Lemut}
 \affiliation{Universit\`a di Genova, Dipartimento di Fisica and INFN Genova,
 	Italy}
\author{R.Menegazzo}
 \affiliation{INFN Padova, Italy }
\author{P.Prati}
 \affiliation{Universit\`a di Genova, Dipartimento di Fisica and INFN Genova,
 	Italy}
\author{V.Roca}
 \affiliation{Universit\`a di Napoli, Dipartimento di Fisica and INFN Napoli, Italy}
\author{C.Rolfs}
 \affiliation{Institut f\"ur Experimentalphysik III,
	Ruhr-Universit\"at Bochum, Bochum, Germany}
\author{M.Romano}
 \affiliation{Universit\`a di Napoli, Dipartimento di Fisica and INFN Napoli, Italy}
\author{C.Rossi Alvarez}
 \affiliation{INFN Padova, Italy }
\author{F.Sch\"{u}mann}
 \affiliation{Institut f\"ur Experimentalphysik III,
	Ruhr-Universit\"at Bochum, Bochum, Germany}
\author{E.Somorjai}
 \affiliation{ATOMKI, Debrecen, Hungary}
\author{O.Straniero}
 \affiliation{Osservatorio Astronomico di Collurania, Teramo and INFN Napoli, Italy}
\author{F.Strieder}
 \affiliation{Institut f\"ur Experimentalphysik III,
	Ruhr-Universit\"at Bochum, Bochum, Germany}
\author{F.Terrasi}
 \affiliation{Seconda Universit\`a di Napoli, Dipartimento di Scienze Ambientali,
	Caserta, and INFN Napoli, Italy}
\author{H.P.Trautvetter}
 \affiliation{Institut f\"ur Experimentalphysik III,
	Ruhr-Universit\"at Bochum, Bochum, Germany}
\author{A.Vomiero}
 \affiliation{Universit\`a di Padova, Dipartimento di Fisica,
	Padova and INFN Legnaro, Italy}
\author{S.Zavatarelli}
 \affiliation{Universit\`a di Genova, Dipartimento di Fisica and INFN Genova,
 	Italy}

\date{\today}

\begin{abstract}

We report on a new measurement of the $^{14}\rm N(\rm p,\gamma)^{15}\rm O$ capture cross section at 
$E_p = 140$ to 400~keV using the 400~kV LUNA accelerator facility at the Laboratori Nazionali del Gran Sasso
(LNGS). The uncertainties 
have been reduced with respect to previous measurements and their analysis. We have 
analyzed the data using the R-matrix method and we find that the ground state transition 
accounts for about 15\% of the total S-factor. The main contribution to the S-factor is given 
by the transition to the 6.79~MeV state. We find a total $S(0) = 1.7\pm0.2$~keV b, in 
agreement with recent extrapolations. The result has important consequences for the solar neutrino
spectrum as well as for the age of globular clusters.

\end{abstract}


\maketitle

\begin{figure}
  \includegraphics[angle=0,width=6cm]{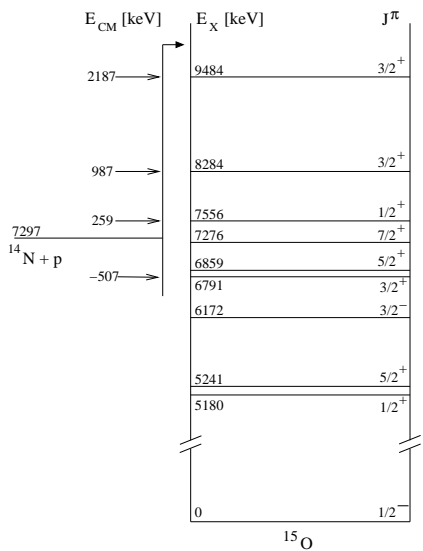}
  \vspace{-0.3cm}
  \caption{\label{fig1}Level structure of $^{15}$O. Above the $^{14}\rm N(\rm p,\gamma)^{15}\rm O$
  threshold only relevant states are shown.}
\end{figure}

The capture reaction $^{14}\rm N(\rm p,\gamma)^{15}\rm O$ ($Q = 7297$~keV) is the slowest process 
in the hydrogen burning CNO cycle \cite{Ro} and thus of high astrophysical interest. This 
reaction plays a role of setting the energy production and neutrino spectrum of
the sun \cite{Ba} as 
well as the age determination of globular clusters \cite{De}. Below 2~MeV, several $^{15}$O states 
contribute to the $^{14}\rm N(\rm p,\gamma)^{15}\rm O$ cross section (fig. \ref{fig1}): a $J^\pi=3/2^+$
subthreshold state at 
$E_R = -507$~keV ($E_x = 6.79$~MeV), and 3 resonant states: $J^\pi=1/2^+$ at $E_R = 
259$~keV, and 3/2$^+$ at $E_R= 987$~keV and  $E_R= 2187$~keV.  The reaction was 
previously studied over a wide range of energies, i.e. $E_{cm} = E = 240$ to 3300~keV (\cite{Sc} and 
references therein). The non resonant capture to excited states in $^{15}$O led the authors of 
ref. \cite{Sc} to an extrapolated astrophysical S-factor at zero energy of $S_{es}(0) = 
1.65$~keV b. The data for the capture into the $^{15}$O ground state were analyzed using 
the Breit-Wigner formalism and indicated an important influence of a subthreshold state at 
$E_R = -507$~keV, leading to $S_{gs}(0) = 1.55\pm0.34$~keV b with a deduced gamma width of 
$\Gamma_\gamma = 6.3$~eV, thus $S_{tot}(0) = 3.20\pm0.54$~keV b.

A reanalysis \cite{An} of the capture data to the ground state \cite{Sc} using an R-matrix approach indicated
a negligible contribution of the subthreshold state to the total S(0)-factor, mainly due to a 
significantly smaller $\Gamma_\gamma$ of this state. This indication was supported by a 
lifetime measurement of the $E_R = -507$~keV subthreshold state via the Doppler-shift 
method \cite{Be} leading to $\Gamma_\gamma = 0.41^{+0.34}_{-0.13}$~eV, and a 
measurement via the Coulomb excitation method \cite{Ya} resulted in $\Gamma_\gamma
 = 0.95^{+0.60}_{-0.95}$~eV. In view of the uncertainties in $S_{gs}(0)$ and thus $S_{tot}(0)$, a new 
measurement of the capture process into the $^{15}$O ground state was highly desirable, 
extending possibly the low energy limit below that of previous work i.e., below $E = 
240$~keV. We report on the results of such measurements using the 400~kV underground 
accelerator facility LUNA at Gran Sasso, Italy. 

\begin{figure*}
  \includegraphics[angle=270,width=16cm]{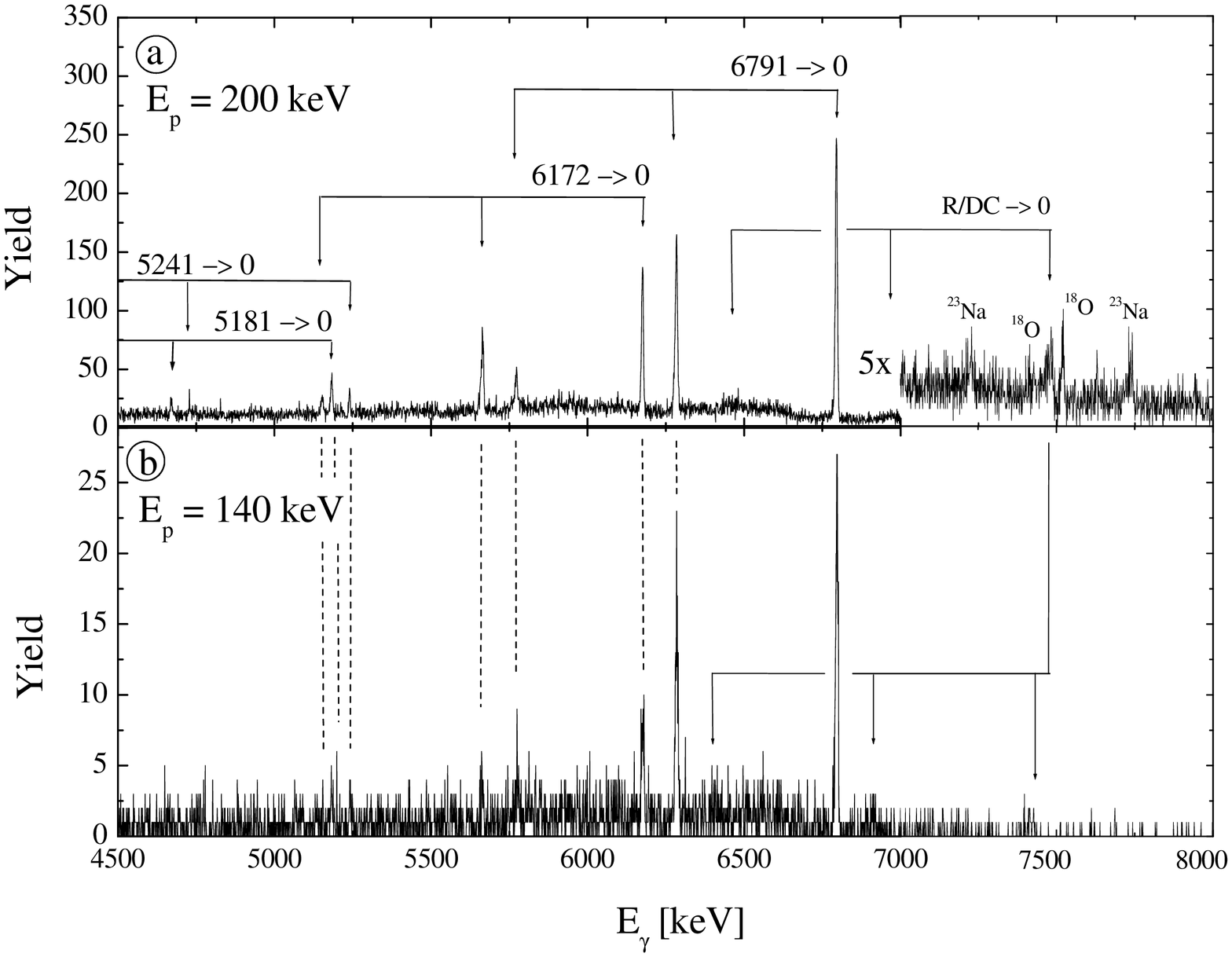}
  \vspace{-0.5cm}
  \caption{\label{fig2}$\gamma$-spectrum for $^{14}\rm N(\rm p,\gamma)^{15}\rm O$
  obtained (a) at $E_p=200$~keV ($\rm Q=85$~C, $\rm t=62$~h) and (b) at
  $E_p=140$~keV ($\rm Q=222$~C, $\rm t=396$~h). For case (a) beam induced background
  originates primarily from traces of $^{11}$B, $^{18}$O, and $^{23}$Na in the target,
  while for case (b) the beam induced background was of the same order than the remaining
  cosmic background which is about 7 counts in the region of the peak for the ground state
  transition where a total of 50 counts are observed.}
\end{figure*}

The 400 kV LUNA facility and the setup have been described elsewhere \cite{Fo}. Briefly, the 
accelerator provided a proton current on target of up to 500~$\mu$A. The absolute energy 
is known with an accuracy of 0.3~keV and the energy spread and the long-term energy 
stability were observed to be 100~eV and 5~eV/h, respectively. Near the target, the beam 
passed a liquid-nitrogen cooled shroud (to minimize carbon-buildup on target) and an 
electrically insulated collimator with a negative voltage of 300~V (to suppress the effects of 
secondary electrons). The water-cooled target was oriented with its normal at 55$^\circ$ to 
the beam direction. The target consisted of a TiN layer (with a typical thickness of 80~keV) 
reactively sputtered on a 0.2~mm thick Ta backing. The target quality was checked frequently 
at the $E_R = 259$~keV resonance: no significant deterioration was observed after a 
bombarding time of several days. Typically, a new TiN target was used after a running time 
of 1 week. The stoichiometry of the TiN layer was verified via Rutherford Backscattering 
Spectrometry using a 2.0~MeV $^4$He$^+$ beam, resulting in $Ti/N=1/(1.08\pm0.05)$.

To deduce cross section values in the non-resonant energy region, one can use either thin targets
(say, a few keV) or thick targets (say, a few 10 keV). Due to sputtering effects at low energies
thin targets deteriorate significantly in a short time changing correspondingly the observed
$\gamma$-ray yields. In the case of thick targets, the sputtering effect on the observed yields
is negligible, as long as the thickness stays large enough, which was verified experimentally
(see above). For the chosen thick targets, we made an analysis of the line shape of the primary
$\gamma$-rays as described below to deduce a capture cross section.

\begin{figure}
  \includegraphics[angle=90,width=9cm]{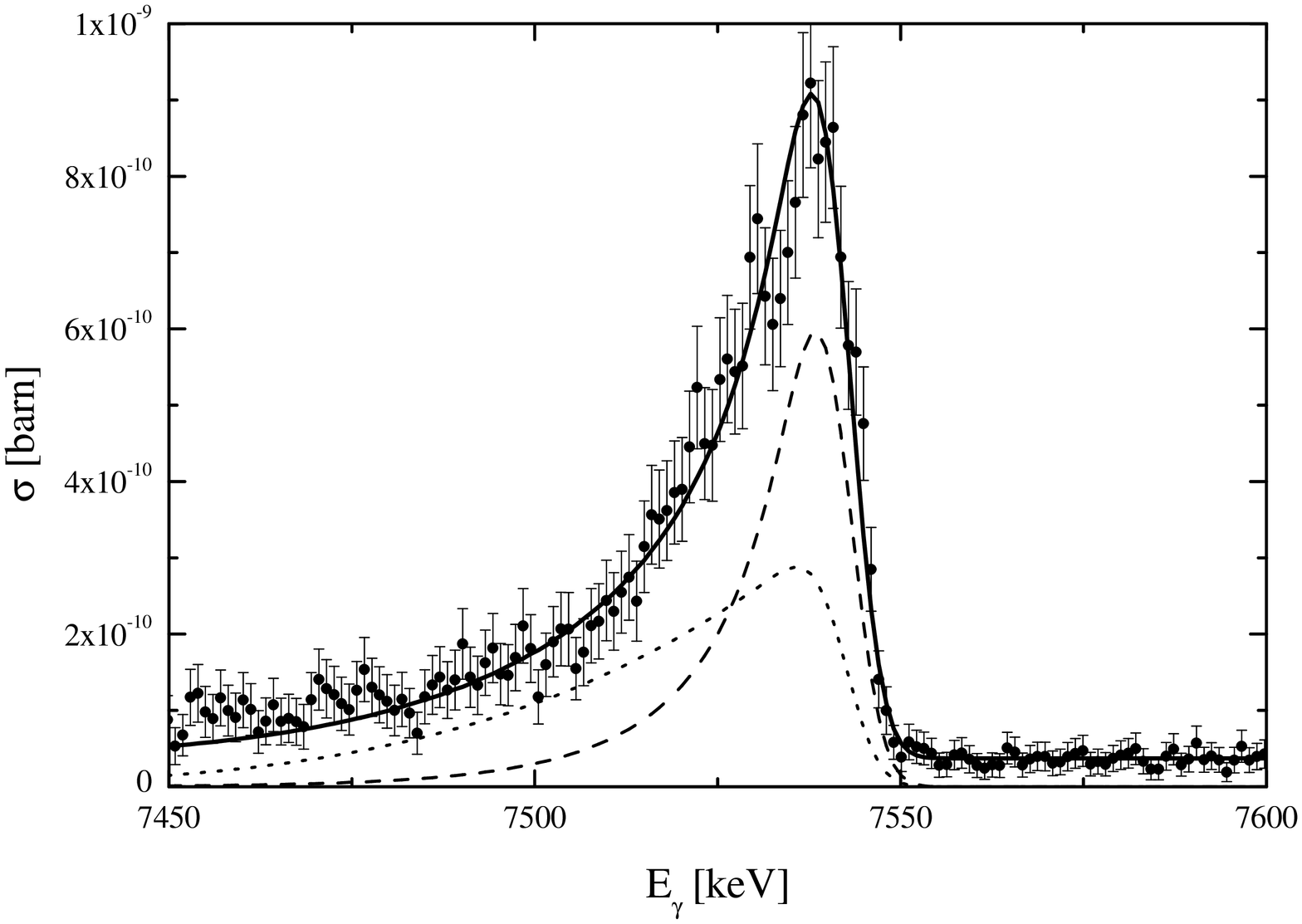}
  \vspace{-1cm}
  \caption{\label{fig3}Typical $\gamma_0$-ray line shape of
  $^{14}\rm N(\rm p,\gamma)^{15}\rm O$ obtained at $E_p=260$~keV. The dashed line
  corresponds to the expected resonant part, the dotted line to the fitted
  non-resonant part, and the solid line is their sum including background.}
\end{figure}

For the measurement of excitation functions the capture $\gamma$-rays were observed with 
one Ge detector (126\% efficiency) placed at  55$^\circ$ in close geometry to the target ($d = 
1.53$~cm was the distance between the target and the front face of the detector). In one 
experiment the distance was increased up to 20.5~cm for the determination of the detector 
efficiency and summing effects. In another experiment, Ge detectors were placed at 0$^\circ$ 
(126~\%), 90$^\circ$  (120~\%), and 125$^\circ$ (108~\%), relative to the beam axis ($d = 
7$~cm) for the measurement of Doppler shifts, excitation energies, and angular distributions.
Sample spectra are shown in fig. \ref{fig2} where the influence of different background sources
can be observed: (a) beam induced background, which is mainly due to proton capture on
$^{11}$B, $^{18}$O, and $^{23}$Na while at $\rm {E_p}=140$~keV (b) the remaning beam induced background
from $^{11}$B contaminations is equal to the cosmic background. 

\begin{table*}
\caption{\label{table1}Excitation energies in $^{15}$O, Doppler shift
measurements and $\gamma$-branching ratios for the 259~keV resonance ($E_x=7556$~keV).}
\begin{ruledtabular}
\begin{tabular}{cccccc}
\hline
\multicolumn{2}{c}{$E_x$ [keV]} & \multicolumn{2}{c}{$F(\tau)$ Doppler shift} 
& \multicolumn{2}{c}{Branching [\%]\footnotemark[1]} \\
present & \cite{Aj} & present\footnotemark[2] & \cite{Be}\footnotemark[3] & present & \cite{Aj} \\
\hline
$ 5180.51\pm0.14$ & $5183.0\pm1.0$ & $0.68\pm0.03$ & $0.68\pm0.03$ & $16.6\pm0.2$
& $15.8\pm0.5$ \\
$ 6171.86\pm0.15$ & $6176.3\pm1.7$ & $0.99\pm0.03$ & $0.91\pm0.05$ & $58.4\pm0.3$
& $57.5\pm0.6$ \\
$ 6791.23\pm0.19$ & $6793.1\pm1.7$ & $0.99\pm0.04$ & $0.93\pm0.03$ & $23.3\pm0.3$
& $23.2\pm0.4$ \\ 
$ 7556.23\pm0.28$ & $7556.5\pm0.4$ & & & $1.67\pm0.10$\footnotemark[4] &
  $3.5\pm0.6$\footnotemark[4] \\
\hline
\end{tabular}
\end{ruledtabular}
\footnotetext[1]{for the $E_R=259$~keV resonance}
\footnotetext[2]{TiN target}
\footnotetext[3]{N implanted in Ta}
\footnotetext[4]{transition to the ground state}
\end{table*}
 
\begin{figure}
  \includegraphics[angle=90,width=9cm]{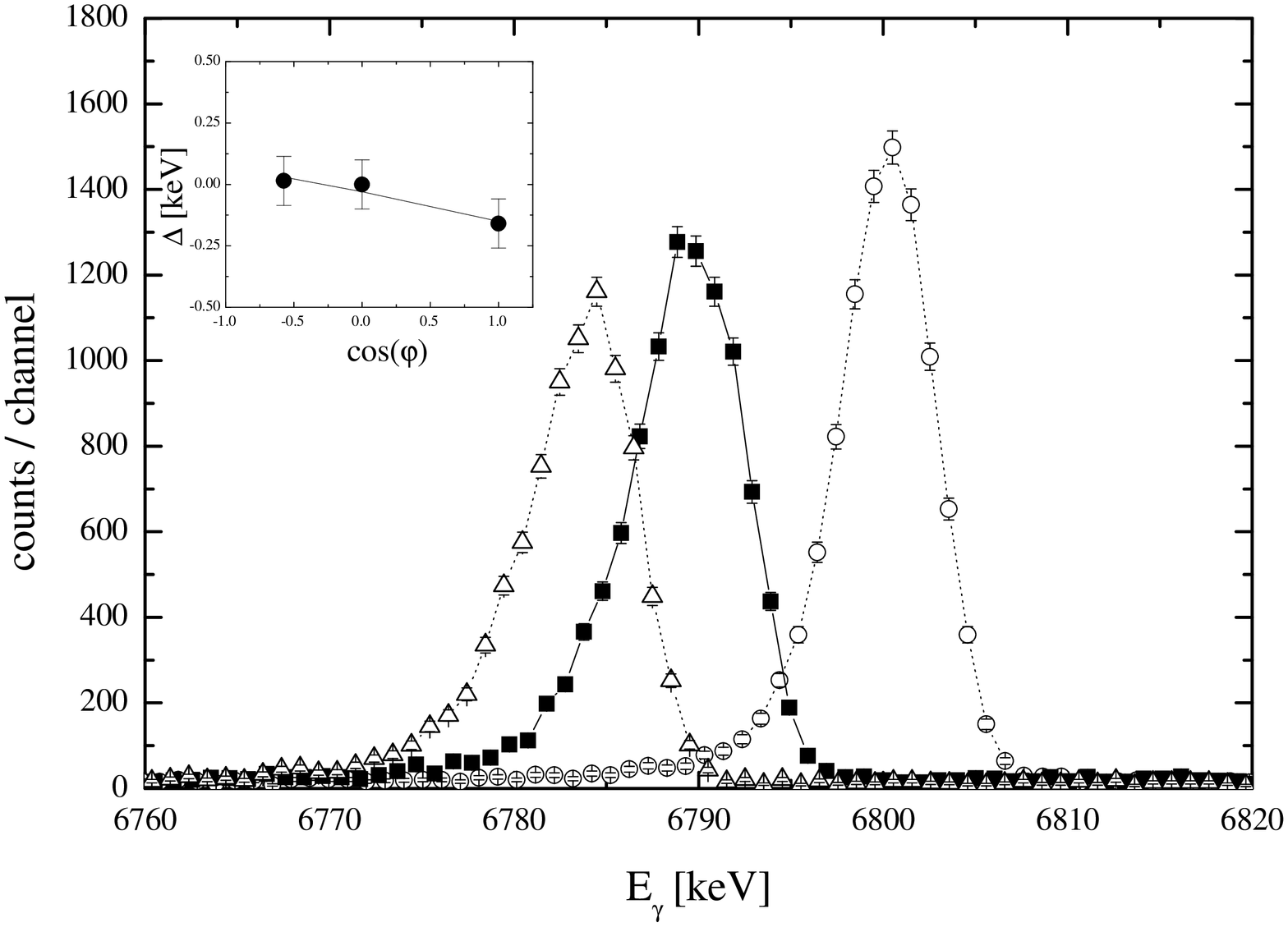}
  \vspace{-1cm}
  \caption{\label{fig4}Full-energy peak for the $6791\rightarrow0$~keV transition
  observed with three Ge detectors positioned at 0$^\circ$ (open circles), 90$^\circ$ (filled squares), and
  125$^\circ$ (open triangles). The insert shows the measured energy shift relative to the
  expected full Doppler shift, where the solid line corresponds to the
  attenuation factor $F(\tau)=0.99$.}
\end{figure}
 
In a similar way as described in \cite{Fo}, the line shape of the $\gamma_0$-capture transition in 
$^{14}\rm N(\rm p,\gamma)^{15}\rm O$ may be used for various purposes. Fig. \ref{fig3} shows the 
$\gamma_0$-ray line shape observed at $\theta_\gamma = 55^\circ$ and $E = 243$~keV, i.e. 
at the low-energy tail of the $E_R = 259$~keV resonance; the dispersion in the spectrum was 
1 keV/channel. The line shape can be interpreted as the sum of the resonant and non resonant 
contributions; when convoluted with the detector resolution, the solid curve through the data 
points is obtained. In this way, the drop in the $\gamma_0$-ray yield towards lower energies 
reflects directly the drop of the cross section. For the calculated line shape the dependence on 
energy of the $\gamma_0$-efficiency and of the stopping power of protons in TiN \cite{Zi} was 
included. The high-energy edge of the peak contains the information on the incident beam 
energy, hence possible C-build up on target could be corrected for; this correction was never 
larger than 2 keV. The method to extract the cross section from the $\gamma_0$-ray line 
shape requires further investigation of the following points: (i) energy, width, branching 
ratios, and strength of the $E_R = 259$~keV resonance, (ii) energy calibration and efficiency 
of the Ge detector, (iii) summing effects in close geometry, (iv) Doppler shift, and (v) effects 
of angular distributions. In the same way the transition to the 6.79~MeV state was analysed. 
Below $E=170$~keV the information of the data from the secondary transition was used in 
the analysis. For this energy region a constant $S_{6.79}$ was assumed over the target 
thickness.  A detailed description of all these points is given elsewhere \cite{Co, Al}, where a 
complete analysis of the data to all final states will be presented.

The resonance energy could in principle be determined from the $\gamma_0$-energy after 
correction for Doppler shift and recoil. One requirement is the precise energy calibration of 
the Ge detector. We have used the information obtained from the experiment with the 3 Ge 
detectors ($\theta_\gamma = 0^\circ$, 90$^\circ$, and 125$^\circ$). From \cite{Be} it is known, that 
the $E_x = 5181$~keV state in $^{15}$O shows an attenuated Doppler shift and the 6172 
and 6791~keV states have lifetimes resulting in nearly full Doppler shifts. Using the known 
energy of the resonance from the accelerator calibration together with 
$Q=7296.8\pm0.5$~keV \cite{Au} and calibration points from radioactive sources we have 
performed a combined $\chi^2$ fit of the data obtained at the three angles. It was necessary to 
vary also the excitation energies of the first three excited states. In addition, we varied the 
attenuation coefficients for the Doppler shifts. The Doppler shift data for the $E_x = 
6791$~keV state are shown in Fig. \ref{fig4} and all results are summarized in Table
\ref{table1}. We confirm 
the results of \cite{Be} for the first excited state in $^{15}$O but could not extract a lifetime for the 
6791~keV state due to a nearly full Doppler shift. For all the subsequent work we calibrated 
the $\gamma$-ray spectra using the new excitation energies given in Table
\ref{table1}.

\begin{figure*}
  \includegraphics[angle=90,width=14cm]{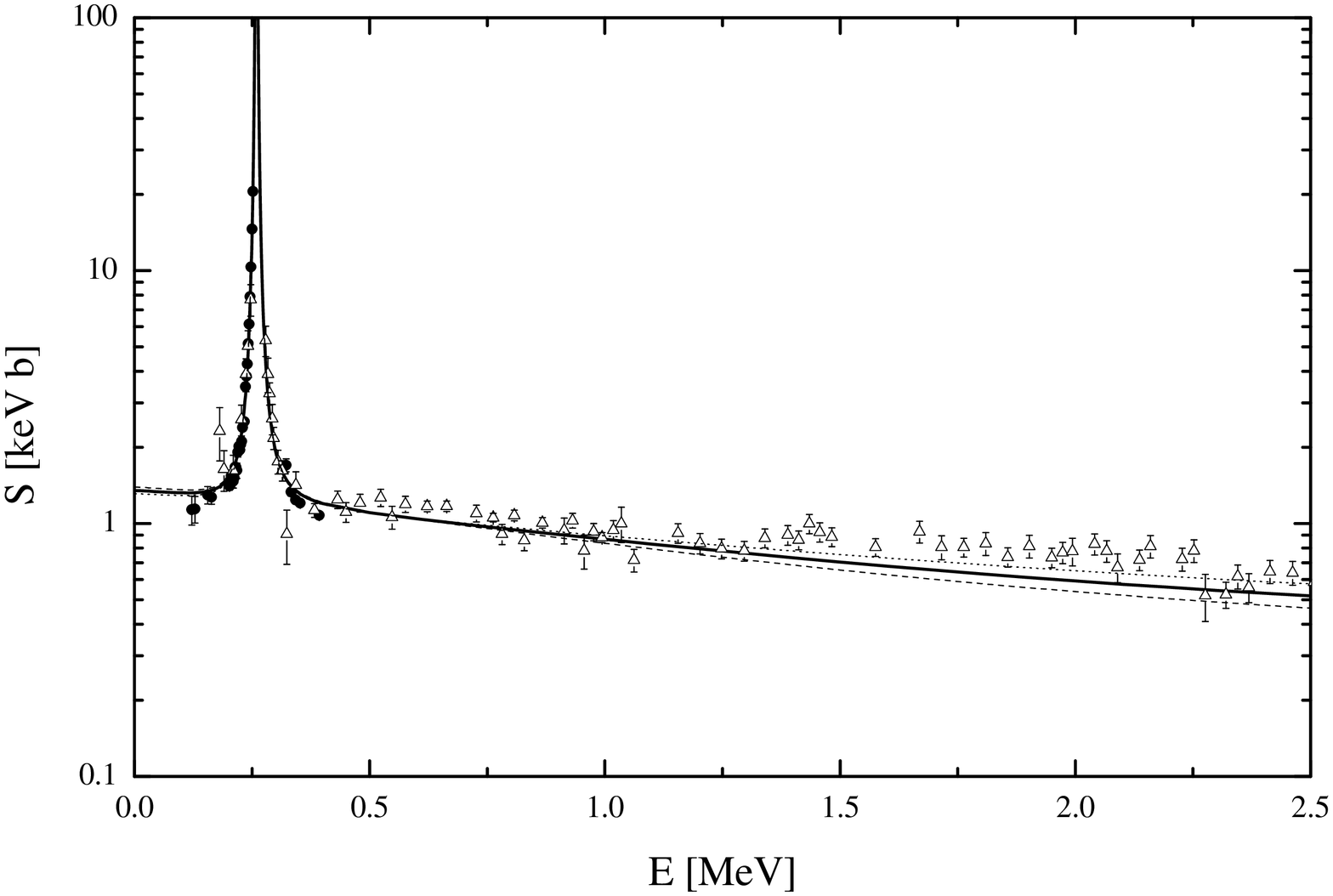}
  \vspace{-0.5cm}
  \caption{\label{fig5}Transition to the 6.79~MeV state in $^{15}$O. The S-factor
  data from present work are represented by solid points, those of
  \cite{Sc} by open triangles. The R-matrix fit (solid line) was obtained for
  $a=5.5$~fm, the dashed line for $a=6$~fm and the dotted line for $a=5$~fm.
  The data point in the 259~keV resonance is off scale.}
\end{figure*}

\begin{table*}
\caption{\label{table2}R-matrix parameters for capture transition to the
ground state}
\begin{ruledtabular}
\begin{tabular}{cccccccc}
\hline
 & ground state & \multicolumn{2}{c}{subthreshold state} &
 \multicolumn{2}{c}{E$_{res}$} & \multicolumn{2}{c}{transition to} \\
 radius & ANC & & & 2.19~MeV & 6~MeV & $E_x=0$~MeV & $E_x=6.79$~MeV \\
 $a$ [fm] & C [fm$^{-1/2}$] & $\gamma^2$ [MeV] & $\Gamma_\gamma$ [eV] & 
 $\Gamma_\gamma$ [eV] & $\Gamma_\gamma$ [eV] & S(0) [keV b] & S(0) [keV b] \\
\hline
 5   & 5.8 & 0.40 & 1.2 & 4.3 & 25 & 0.19 & 1.31 \\
 5.5 & 7.3 & 0.42 & 0.8 & 4.4 & 23 & 0.25 & 1.35 \\
 6   & 8.8 & 0.44 & 0.5 & 4.7 & 26 & 0.31 & 1.39 \\
\hline
 average & & $ 0.42\pm0.02$ & $0.8\pm0.4$ & $4.5\pm0.2$ & $23\pm3$ &
 $0.25\pm0.06$ & $1.35\pm0.05$ \\
\hline
\end{tabular}
\end{ruledtabular}
\end{table*}

The detector efficiency was determined using calibrated radioactive sources and the cascade 
condition for the transitions to the first three excited states at the $E_R = 259$~keV 
resonance, as for all of them no other decay than that to the ground state was observed. This procedure 
was performed with the Ge detector placed at 1.53, 5.5, 10.5, and 20.5~cm distances from the 
target in order to determine the summing-in contribution to the ground state transition and the 
summing-out for the transitions to the excited states. It turned out, that the
summing-in yield 
was about 3.5 times higher than the actual ground state intensity at the 1.53~cm position. This 
1.53 cm position was used for the entire cross section measurements. The efficiency curve 
was also calculated using the GEANT routine \cite{Geant} and found to be 
in excellent agreement with observation \cite{Co}. Branching and strength values of the $E_R= 
259$~keV resonance were determined with low beam current to avoid dead time effects and 
at far distance (20.5~cm) to minimize summing effects from cascade transitions. The 
branching results are given in Table \ref{table1}; they are in good agreement with previous work \cite{He}, 
except for the ground state transition. Our strength value of $\omega\gamma = 13.5\pm0.4$ 
(statistical) $\pm0.8$(systematical) meV, which was determined absolutely,
is also in good agreement with previous work \cite{Sc, Aj}, 
$\omega\gamma = 14\pm1$~meV. The fitted total width of $\Gamma=0.99\pm0.03$~keV is consistent
with previous work (\cite{Sc}: $\Gamma=1.2\pm0.2$~keV).

All transitions show isotropy at the $E_R = 259$~keV, 
$J^\pi=1/2^+$ resonance and no forward-backward asymmetry \cite{Co} outside the resonance. 
Consequently, an angle-integrated $\gamma_0$-ray yield may be derived directly from the 
yield collected by the Ge detector at $\theta_\gamma= 55^\circ$. Finally, the summing due to 
the large solid angle of the detector and the cascade coincident events in the transition to the 
ground state was corrected for by studying all other capture transitions. The absolute scale of
the cross section deduced from the $\gamma$-ray line shape was obtained by normalizing to the
thick target yield in the 259 keV resonance \cite{Ro} whose strength $\omega\gamma$ was determined
absolutely (see above).

\begin{figure*}
  \includegraphics[angle=90,width=14cm]{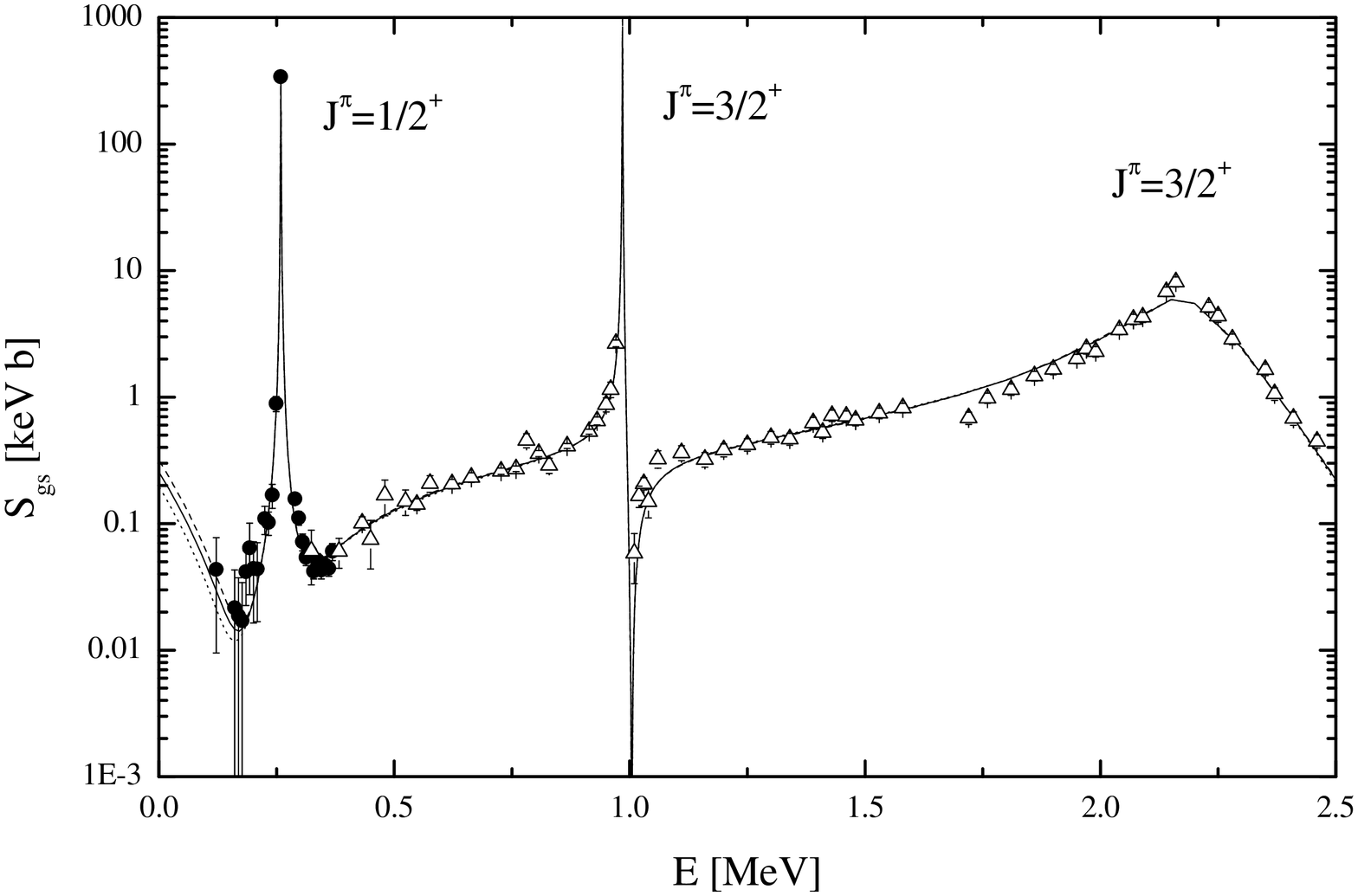}
  \vspace{-0.5cm}
  \caption{\label{fig6}Astrophysical S(E)-factor curve for the ground state
  transition in $^{14}\rm N(\rm p,\gamma)^{15}\rm O$. Filled-in data points are the results
  from LUNA, while the open data points are from previous work \cite{Sc} correct
  for summing effects. The solid line corresponds to the R-matrix fit for
  $a=5.5$~fm, the dashed line for $a=6$~fm and the dotted line for $a=5$~fm.}
\end{figure*}

The capture data from LUNA and previous work \cite{Sc} for the transition to the 6.79~MeV
state are shown in figure \ref{fig5}: there is an excellent agreement between both data sets
in the region of overlap. As shown from previous work \cite{Sc, An}, the S-factor has an external
non-resonant contribution and a contribution from the $\rm{E_R}=259$~keV resonance, where the
latter contribution was derived from $\omega\gamma$ and $\Gamma$ values of the present work.
The R-matrix fit \cite{An, La} included both contributions. The resulting ANC value of
$C = 5.0\pm0.1$~fm$^{-1/2}$ agrees with indirect determinations (\cite{Mu}: 
$C=5.2\pm0.7$~fm$^{-1/2}$; \cite{Ber}: $C = 4.6\pm0.5$~fm$^{-1/2}$) and leads to the fitted
curve shown in figure \ref{fig5}. For large values of the R-matrix radius $a$ the fit deviates
from the data for energies above $\rm E=1$~MeV (see Fig. \ref{fig5}). We have adopted the results for
$a=5.5$~fm leading to the extrapolated value $\rm S_{6.79}=1.35\pm0.05$ (statistical) $\pm0.08$
(systematical)~keV b. This value is about 20~\% lower than the R-matrix fit \cite{An} of the data
from \cite{Sc} alone.

The data from LUNA and previous work \cite{Sc} for the ground state capture are shown in figure
\ref{fig6}. We have corrected the previous data for summing effects, which are at most
10~\% above $\rm E=500$~keV, except near the destructive interference structure of the
${\rm E_R}=987$~keV resonance. The data were fitted including
the 3/2$^+$ subthreshold state, the 1/2$^+$, 259~keV, the 3/2$^+$, 987~keV and the 
3/2$^+$, 2187~keV resonances as well as a background pole located at 6~MeV. For the 
subthreshold state, we used the reduced width $\gamma^2$ obtained from the fit of the data 
for the 6.79~MeV transition. The fit parameters were the $\Gamma_\gamma$ of the 
subthreshold state, the 987~keV and 2187~keV resonances, the $\Gamma_p$ and 
$\Gamma_\gamma$ of the background pole. For the external contribution we used the ANC 
of \cite{Mu} as a starting value. The results are shown in Fig. \ref{fig6} for $a =5$~fm
and $C_{gs} = 5.8$~fm$^{-1/2}$
 (dotted line), $a=5.5$~fm and $C_{gs} = 7.3$~fm$^{-1/2}$ (solide line), and $a =6.5$~fm 
and $C_{gs} = 8.8$~fm$^{-1/2}$  (dashed line). The corresponding 
$\Gamma_\gamma$ values for the subthreshold state and extrapolated S-factor values are 
listed in Table \ref{table2}.  It can be noted that our deduced value $\Gamma_\gamma =0.8\pm0.4$~eV 
is in good agreement with  the value from a life time measurement by \cite{Be} 
$\Gamma_\gamma=0.41^{+0.34}_{-0.13}$~eV as well as  with 
$\Gamma_\gamma=0.95^{+0.60}_{-0.95}$~eV, the value from coulomb excitation work
\cite{Ya}. 
All $\Gamma_\gamma$ values from Table \ref{table2} are consistent within uncertainties with our 
observation of a nearly full Doppler shift for the transition of the 6.79~MeV state
(see Fig. \ref{fig4}). The 
best $\chi^2$ for the fit to the ground state transition is obtained for $a =6$~fm ($\chi^2 = 
1.27$), however the differences are small (for $a =5$~fm, $\chi^2=1.35$; contributions to 
$\chi^2$ near the narrow resonances have been omitted in the $\chi^2$ values given above). 
The best overall agreement with all available data \cite{Sc, Be, Ya, Ber, Mu} is obtained for $a=5.5$~fm, 
where our fitted ground state ANC is $C=7.3$~fm$^{-1/2}$, that can be compared with the
result of \cite{Mu} 
which gives $C = 7.3\pm0.4$~fm$^{-1/2}$, after conversion to the coupling scheme used in 
the present work. For the total S-factor a contribution from the transition to
the 6.17~MeV 
state of  $S_{6.17}(0)=0.06$~keV b from \cite{An} has been added to obtain an average ($a=5$ to 
6~fm) of  $S_{tot}(0)=1.7\pm0.1$~(statistical) $\pm0.2$(systematical) keV b, which can be 
compared with $1.77\pm0.20$~keV b \cite{An} and $1.70\pm0.22$~keV b \cite{Mu} from previous 
analyses.

    In summary, the present work improves the experimental information concerning the 
ground state transition in $^{14}\rm N(\rm p,\gamma)^{15}\rm O$. The previous data \cite{Sc} corrected by 
summing effects discussed above are now in good agreement with the present work. The large 
ambiguity for the extrapolation of the ground state transition has been reduced. In the energy 
region between $E=170$ to 300~keV the S(E)-factor is dominated by the 259~keV resonance. 
Only at energies of about 100~keV and below the contributions of the subthreshold resonance 
and non-resonant mechanisms should become clearly observable. Using large volume Ge 
detectors it is impossible to obtain additional information on S$_{gs}$(E) at lower energies 
(than current work) with acceptable uncertainty due to the sizable summing correction. An 
improved information on S$_{tot}$(E) can be achieved possibly using a 4$\pi$-summing 
crystal to observe the total S(E)-factor. Such an experiment is presently underway at the 
LUNA facility.

After submission of the present work several relevant publications appeared:

(i) The authors of \cite{Ne} reanalyzed the data of \cite{Sc} based on
conclusions from their analyzing power data. This led to a different
extrapolation for the transition to the 6.79~MeV state (about 10 to 20~\% lower
than the analysis of \cite{An} and 11~\% higher than present work) and for the transition to the
6.17~MeV state (about a factor 3 higher than \cite{An}). However, the primary
aim of the present letter was to clarify the strength of the ground state
transition.

(ii) The analysis in \cite{Im, Fi, Ba04} were partly based on the results of the
present work and led to the conclusion that the solar neutrino flux from CNO
cycle is reduced by a factor 2 and the age of the globular clusters is increased
by about 1 billion years.

\bibliography{luna}

\end{document}